# Spin caloritronics: History and future prospects of experiments

Ken-ichi Uchida [a,b,*], Takamasa Hirai [a]

[a] Research Center for Magnetic and Spintronic Materials, National Institute for Materials Science, Tsukuba 305-0047, Japan

[b] Department of Advanced Materials Science, Graduate School of Frontier Sciences, The University of Tokyo, Kashiwa 277-8561, Japan

[*] UCHIDA.Kenichi@nims.go.jp

**Abstract**

Since the beginning of the 21st century, novel energy conversion and control principles utilizing the spin degree of freedom have been discovered in the field of spin caloritronics, which integrates spintronics with thermal transport and thermoelectric properties. In this article, we review the history of development of spin caloritronics and experimental studies on various transport phenomena caused by heat-charge-spin interactions. We then discuss future prospects in spin caloritronics from the viewpoints of measurement techniques, physics, materials science, and engineering applications. Spin caloritronics is now at a turning point, transitioning from fundamental condensed matter physics to materials science, and further development is anticipated in both fundamental and applied research.

## 1. Introduction

Spin caloritronics has evolved by introducing heat into spintronics. The origin of spin caloritronics is traced back to the nonequilibrium thermodynamics theory concerning the heat-charge-magnetization interaction in magnetic heterostructures reported by Johnson and Silsbee in 1987 [1]. However, since the concept of a spin current, a flow of spin angular momentum, was not yet established at this time and experimental verification was slow to progress, physics and functionalities appearing due to heat-spin interactions remained unexploited in spintronics for a long period. The term "spin caloritronics" was first used historically in 2007 by Hatami et al. [2]. They theoretically showed that the magnetization of a ferromagnet could be reversed by a spin-transfer torque induced by a heat flow. The following year, Uchida et al. experimentally observed a spin current generated by applying a temperature gradient to a ferromagnetic metal and named the heat-to-spin current conversion phenomenon the spin Seebeck effect (SSE) (Fig. 1) [3]. SSE experiments were distinct from other spintronics research, performed on simple magnetic bilayer film structures requiring no microfabrication. The low barrier to entry for this experimental approach led to rapid progress in verifying SSE and elucidating its principles. Consequently, spin caloritronics rapidly grew into an interdisciplinary field studied worldwide. In this field, since the discovery of SSE, novel magneto-thermoelectric and thermo-spin conversion effects have been constantly found almost every year and the international workshops on spin caloritronics have been held worldwide (Fig. 1).

The development of spin caloritronics has had several turning points. Even today, by incorporating new insights, technologies, and materials, it continues to achieve unique progress in terms of measurement techniques, condensed matter physics, materials science, and engineering applications. This review summarizes the fundamental classification and properties of various spin-caloritronic phenomena, i.e., transport phenomena arising from heat-charge-spin interactions, provides an overview of the entire field, and discusses recent progress and perspective on future directions. Due to space constraints, it is not possible to delve deeply into individual topics. However, it is hoped that this review will serve as a resource for surveying research trends in this field and as a catalyst for new interdisciplinary collaborations.

## 2. Classification of spin-caloritronic phenomena

While various approaches exist for classifying spin-caloritronic phenomena, this review broadly categorizes them into three types: (1) magneto-thermoelectric effects, i.e., thermoelectric effects dependent on magnetic fields or magnetization, (2) thermomagnetic effects, i.e., thermal transport dependent on magnetic fields or magnetization, and (3) thermo-spin effects, i.e., conversion between heat and spin currents (Fig. 2) [4]. They are further classified into longitudinal effects, where the output current is parallel to the input current, and transverse effects, where the output current is perpendicular to the input current (Fig. 2). Hereafter, for example, magneto-thermoelectric effects in which the input current is parallel (perpendicular) to the output current are denoted as longitudinal (transverse) magneto-thermoelectric effects. Spin currents are transported by various quasiparticles possessing spin angular momentum, but this review focuses specifically on spin currents carried by conduction electrons and magnons, which have been the primary subjects of research in spin caloritronics.

### *2.1. Magneto-thermoelectric effects*

In addition to the conventional Seebeck and Peltier effects, applying a magnetic field to metals or semiconductors manifests diverse magneto-thermoelectric effects (Fig. 2). Representative examples include the magneto-Seebeck (Peltier) effect, where the Seebeck (Peltier) coefficient changes depending on an external magnetic field; the Nernst effect, where a thermopower is generated in the direction of the cross product of the applied magnetic field and temperature gradient; and the Ettingshausen effect, where a heat current is generated in the direction of the cross product of the applied magnetic field and charge current. Here, the magneto-Peltier and Ettingshausen effects are the Onsager reciprocals of the magneto-Seebeck and Nernst effects, respectively. These magneto-thermoelectric effects are significantly pronounced in specific materials such as Bi and Bi-based alloys. Applied research, such as electronic cooling using the Ettingshausen effect, has been conducted for a long time [5,6]. Although these phenomena have not yet been commercialized due to the requirement for strong external magnetic fields, detailed mechanism elucidation and new material development continue to progress [7]. On the other hand, spin caloritronics primarily targets the magneto-thermoelectric effects unique to magnetic materials with spontaneous magnetization. In magnetic conductors, various magneto-thermoelectric effects dependent on magnetization manifest even without an external magnetic field by acting the spin-orbit interaction on conduction electrons (note that to distinguish the magnetic-field-dependent effects from the magnetization-dependent effects, the former effects are widely called the ordinary Nernst and Ettingshausen effects, while the latter effects are the anomalous Nernst and Ettingshausen effects). Despite the fact that the existence of many magneto-thermoelectric phenomena has been known for a long time, they have not been sufficiently studied. They remain attractive research subjects with numerous challenges in both fundamental and applied aspects.

The longitudinal magneto-thermoelectric effect in magnetic materials is called the anisotropic magneto-Seebeck effect (Fig. 2). It is the thermoelectric counterpart of the anisotropic magnetoresistance; the Seebeck coefficient changes anisotropically depending on the relative angle between the applied temperature gradient and magnetization [8]. Due to the reciprocity of transport phenomena, the anisotropic magneto-Peltier effect (AMPE) also exists, where the Peltier coefficient varies depending on the relative angle between the charge current and magnetization. The anisotropic magneto-Seebeck effect has been observed in various materials and contributions originating from the anisotropy of the Seebeck and Peltier coefficients were indirectly observed in spin transport experiments on magnetic nanostructures [9]. Consequently, the existence of AMPE has been treated as a matter of course. However, the temperature change induced solely by AMPE was not observed directly before the active thermal imaging technique called the lock-in thermography was introduced into spin caloritronics [10]. The direct observation of AMPE led to the successive discovery of novel thermal control functionalities unachievable without utilizing spin or magnetism [4]. These include electronic cooling in a single material without heterojunctions and thermoelectric conversion where heat absorption/release locations can be freely designed and reconfigured by altering magnetization distribution and device geometry. This development exemplifies how advancements in thermal measurement techniques played a crucial role in the progress of spin caloritronics (see Section 3.1).

A representative example of the transverse magneto-thermoelectric effect in magnetic materials is the anomalous Nernst effect (ANE) [Fig. 3(a)], where a thermopower is generated in the direction of the cross product of the temperature gradient and magnetization [11,12]. Fundamental research on ANE has been conducted for a long time, but it has gained momentum with the development of spin caloritronics [Fig. 3(c)]. One of the triggers to boost the ANE studies during the dawn of spin caloritronics was the experimental activity aimed to separate the contribution of SSE from ANE, which was necessary to firmly establish the existence of SSE [note that the thermopower due to ANE and SSE exhibits similar symmetry, as shown in Figs. 3(a) and (b)] [13]. Since around 2017, topological materials began to be introduced into spin caloritronics research [14]. Following the observation of giant ANE in topological ferromagnet $Co_2MnGa$ in 2018, material development and fundamental understanding on ANE have accelerated further, becoming a trend in condensed matter physics [15]. Regarding the anomalous Ettingshausen effect (AEE), the Onsager reciprocal of ANE, experiments using several typical bulk ferromagnetic metals were reported in the 1920s [16,17], but little research followed thereafter. It was not until 2018 that AEE became observable not only in bulk materials but also in thin-film devices, leading to the emergence of experimental reports on various materials [18]. Then, significantly large AEE has been observed in practical permanent magnets such as $SmCo_5$-type magnets [19]. The transverse magneto-thermoelectric conversion due to ANE and AEE exhibits convenient scaling, where the output power is proportional to material size. This holds potential for realizing simple, versatile thermoelectric applications. In fact, research and development are progressing toward implementing energy harvesting devices [20] and heat flux sensors [21,22].

*2.2 Thermomagnetic effects*

The longitudinal and transverse thermomagnetic effects are referred to as the magneto-thermal resistance (MTR) and thermal Hall effect, respectively, that can serve as principles for magnetic thermal switching. Although these phenomena have also been known for a long time, research examples are fewer compared to the magneto-thermoelectric effects and thermo-spin effects discussed later.

MTR refers to the magnetic field or magnetization dependence of thermal conductivity and is often called the Maggi-Righi-Leduc effect. The magnetization direction dependence of thermal conductivity in ferromagnetic metals, i.e., anisotropic MTR, is typically very small (less than a few percent). In contrast, MTR in magnetic multilayer films, where ferromagnetic metals and nonmagnetic materials are alternately stacked at the nanoscale, has potential to serve as a fundamental principle for active thermal management technologies. As is widely known, such multilayer films form the core of spintronic devices and exhibit the giant magnetoresistance when the nonmagnetic layer is metallic and the tunnel magnetoresistance effect when it is an insulating tunnel barrier, where the electrical conductivity is higher when the magnetization directions of adjacent ferromagnetic layers are parallel than when they are antiparallel. In such magnetic multilayer films, not only the electrical conductivity but also the thermal conductivity depends on the magnetization configuration. Recent experiments have developed the epitaxial Cu/CoFe multilayer film where the thermal conductivity switching ratio reaches 150% at room temperature [23]. Remarkably, this thermal conductivity switching ratio is significantly larger than the magnetoresistance ratio, i.e., the electrical conductivity switching ratio, and its origin remains unexplained [24]. Since MTR in magnetic multilayer films can be directly applied to thermal management for spintronic devices, further improvement of the thermal conductivity switching performance and elucidation of its microscopic mechanism are crucial.

Recent studies have highlighted the potential of magnons for controlling the longitudinal thermal transport in ferromagnetic metals [25,26]. In 2025, it has been reported that the transport of a nonequilibrium magnon spin current dependent on its boundary condition significantly modulates thermal conduction in a junction comprising a ferromagnetic metal thin film and nonmagnetic or magnetic insulator even at room temperature [27]. Such boundary-dependent magnon transports are intimately connected to the spin diffusion model investigated in SSE studies [28], wherein spin currents are blocked at a ferromagnetic metal/nonmagnetic insulator interface [Fig. 4(a)] but transmitted at a ferromagnetic metal/magnetic insulator interface [Fig. 4(b)]. Hirai et al. experimentally proved that the difference in the boundary condition for spin currents changes the thermal transport properties; using CoFe and NiFe thin films as ferromagnetic metals, gadolinium gallium garnet $Gd_3Ga_5O_{12}$ (GGG) as a nonmagnetic insulator, and yttrium iron garnet $Y_3Fe_5O_{12}$ (YIG) as a magnetic insulator, the thermal conductivity of the ferromagnetic metals and the interfacial thermal conductance with the adjacent insulator are modulated by ~10 $Wm^{-1}K^{-1}$ and ~0.5 $GWm^{-2}K^{-1}$, respectively, depending on whether the insulator is GGG or YIG [Figs. 4(c) and (d)]. It is noteworthy that, although the scenario based on the spin diffusion model is applicable to both conduction-electron and magnon spin currents, the observed change of the thermal transport properties was found to originate from magnon spin currents by examining the composition dependence of the ferromagnetic metal and by inserting a nonmagnetic metal layer between the ferromagnetic metal and insulator. The observed large change clearly demonstrates the significant role of magnons in nonequilibrium thermal conduction. Conventional thermal engineering has predominantly centered on controlling the phonon transport through strategies such as phonon scattering and defect engineering to tailor thermal conductivity, i.e., phonon engineering [29], while contributions from other quasiparticles have been widely regarded as insignificant at room temperature. The experimental results in Ref. [27] go beyond this long-standing wisdom and open another principle for thermal transport engineering based on magnons, termed magnon engineering. This emerging framework provides a new engineering opportunity for thermal management technologies based on spin caloritronics.

The thermal Hall effect is the thermal counterpart of the Hall effect and is also known as the Righi-Leduc effect. When the thermal Hall effect manifests, a heat current bends perpendicular to an external magnetic field or magnetization; reversing the magnetic field or magnetization inverts the direction of the transverse heat current. In a similar manner to the Hall effect, the component dependent on the magnetic field (magnetization) is often referred to as the ordinary (anomalous) thermal Hall effect. The thermal Hall effect manifests not only in conductors but also in insulators where phonons and/or magnons are heat carriers [30,31]. The thermal Hall effect has also been used in recent years as a probe to verify the existence of Majorana fermions in magnetic insulators, although it is not directly related to spin caloritronics [32].

### *2.3. Thermo-spin effects*

While the magneto-thermoelectric effects discussed in Section 2.1 are spin-dependent phenomena, they are the conversion phenomena between heat and charge currents. Conversely, in spin caloritronics, conversion phenomena between heat and spin currents have been successively discovered (Figs. 1 and 2). These thermo-spin effects can be broadly categorized into the conversion between a heat current and conduction-electron spin current and the conversion between a heat current and magnon spin current. The former arises from the spin dependence of the Seebeck coefficient, i.e., the difference in the Seebeck coefficient between up-spin and down-spin electrons. The phenomenon where a conduction-electron spin current is generated from a temperature gradient is called the spin-dependent Seebeck effect. It was first observed experimentally by Slachter et al. in 2010 [33]. They used a nonlocal approach in an in-plane spin-valve structure to observe spin accumulation induced by a temperature difference at a ferromagnetic metal/paramagnetic metal interface. In 2011, Le Breton et al. demonstrated that spins can be injected into Si via a tunnel junction using a ferromagnetic metal/oxide/Si junction system [34]. In 2012, Flipse et al. observed the spin-dependent Peltier effect, the reciprocal of the spin-dependent Seebeck effect, using a nanopillar-type spin-valve device with a micro-thermocouple [35]. Using similar devices, the observation of a temperature difference between spin channels, i.e., spin heat accumulation, was also reported [36]. These heat-spin current conversion phenomena occur exclusively in itinerant electron systems, and are fundamentally different in origin from SSE and the spin Peltier effect (SPE) mediated by a magnon spin current.

The discovery of SSE became the catalyst for the explosive development of spin caloritronics. SSE is a phenomenon where a temperature gradient applied to a system consisting of a magnetic material and conductor induces a spin current near the junction interface. To detect the spin current generated by SSE, the inverse spin Hall effect [37,38] in a metallic film formed on the magnetic material is widely used. By fabricating a metallic film with strong spin-orbit interaction, such as Pt, the spin current can be converted into an electric voltage signal, enabling the electrical detection of SSE [Fig. 3(b)]. In the first experiments on SSE in 2008, a junction between a ferromagnetic metal ($Ni_{81}Fe_{19}$) and Pt was used [3]. Consequently, SSE was believed to be driven by a conduction-electron spin current and to occur only in metals or semiconductors. In other words, initial discussions of SSE were based on a mechanism similar to the spin-dependent Seebeck effect described above. Amidst this, experiments reported in 2010 demonstrated that SSE also manifests in magnetic insulators, such as YIG, where conduction-electrons' transport is forbidden [39,40]. This fact provides clear evidence that SSE is driven by magnons, which can transfer spin angular momentum even in magnetic insulators [41,42]. Since then, both experimental and theoretical studies on SSE have advanced dramatically. Experiments using magnetic insulators have become mainstream due to the relatively high heat-spin current conversion efficiency achievable through long-range magnon transport and the ease of separating the spin-current effects from the magneto-thermoelectric effects.

Considering the difference in driving carriers between SSE and the spin-dependent Seebeck effect, it might seem natural to refer to the former as the magnon Seebeck effect and the latter as the spin Seebeck effect. However, by 2010 when the direct observation of the spin-dependent Seebeck effect was reported, the term "spin Seebeck effect" and its experimental methodology had already become globally established to discuss the heat-spin current conversion in macroscale devices. Consequently, the phenomenon driven by magnons (conduction electrons) came to be called SSE (the spin-dependent Seebeck effect). Although this nomenclature in the thermo-spin effects may be complicated due to historical reasons, SSE and the spin-dependent Seebeck effect are entirely distinct phenomena. The same applies to the corresponding reciprocal phenomena: SPE and the spin-dependent Peltier effect.

SPE was first observed in 2014 by Flipse et al. using an YIG/Pt junction [43]. Subsequently, in 2016, Daimon et al. established a measurement method for SPE based on the lock-in thermography [44]. This enabled the direct visualization of temperature changes induced by spin currents in simple structures that do not require microfabrication processes. This experiment revealed that spin currents are accompanied by unique temperature changes that do not broaden spatially.

The YIG/Pt junction system has served as a platform for elucidating the physics of SSE, SPE, and the magnon-electron spin transport at magnetic insulator/metal interfaces. Systematic studies reported during the 2010s largely clarified these fundamental mechanisms. While SSE and SPE are driven by exchange magnons, the measurements for the YIG/Pt systems under high magnetic fields at low temperatures have revealed that subthermal magnons, the frequency range of which is much lower than that where the magnon number is maximized in thermal equilibrium, dominate the heat-spin current conversion [45,46]. The spin-current detection technique using the inverse spin Hall effect in metallic films has become standard methods for investigating the principles and functionalities of the thermo-spin effects. By using the technique and by replacing YIG with various materials, it has been confirmed that SSE occurs also in paramagnets [47], antiferromagnets [48], and multiferroic materials [49], revealing that SSE is a universal phenomenon present in any magnetic material. It has also been shown that not only conduction electrons and magnons, but also spinons [50], triplons [51], nuclear spins [52], and hybridized quasiparticles such as magnon-polarons [53] can be the origins of the thermo-spin effects.

The thermo-spin effect driven by magnetostatic spin waves also exists (Fig. 2). In 2013, An et al. discovered that, when nonreciprocal surface spin waves were excited in YIG via microwave-induced magnetic resonance, heat was transferred unidirectionally, with heating occurring at positions several millimeters or more away from the microwave irradiation point [54]. Due to the nonreciprocal nature of surface spin waves, the direction of heat transport reverses with the magnetization reversal of YIG. This phenomenon, termed the unidirectional spin-wave heat conveyer, originates from the magnetic dipole-dipole interaction, the length scale of which is several orders of magnitude greater than that of exchange magnons. Consequently, novel thermal control functionalities have been demonstrated, such as generating heat via spin waves across spatial gaps [55].

## 3. Prospects for spin caloritronics

Building on the background outlined in Section 2 and our recent research progress, this section presents an outlook for the future of spin caloritronics, considering aspects of measurement techniques, fundamental physics, materials science, and engineering applications.

### *3.1. Measurement techniques*

Advances in thermal measurement techniques have brought novel developments to spin caloritronics. Early spin caloritronics research focused on phenomena generating a spin or charge current from a heat current, such as SSE and ANE, with an eye toward future applications in thermoelectric generation technology. The turning point in this situation was the realization of thermal imaging of SPE via the lock-in thermography technique [44]. The lock-in thermography, one of the active infrared emission microscopy techniques, provides a much better signal-to-noise ratio and higher sensitivity than conventional thermography. During the lock-in thermography measurements, one applies a periodic external perturbation, e.g., a charge current, heat current, microwave, magnetic field, light, and strain, to a sample and extract the first harmonic response of thermal images oscillating with the same frequency as the perturbation [56,57] (note that multi-harmonic lock-in thermography is also possible [58]). The obtained thermal images are converted into the lock-in amplitude and phase images by Fourier analysis. Here, the amplitude image shows the distribution of the magnitude of temperature modulation induced by the external perturbation, and the phase image provides information about the sign of temperature change and time delay due to thermal diffusion. If the input perturbation is a square-wave-modulated AC charge current with zero offset, one can visualize the temperature modulation induced by thermoelectric effects free from Joule heating background in the first harmonic response. Through the spin-orbit interaction, the same method is applicable to the measurements of the thermo-spin effects (Fig. 5). The establishment of the lock-in thermography measurement and analysis for spin-caloritronic phenomena has led not only to the visualization of SPE but also to the observation of AEE in thin films [18] and permanent magnets [19], AMPE [10], the magneto-Thomson effect in a nonmagnetic conductor [59], anisotropic magneto-Thomson effect in a ferromagnetic metal [60], and the transverse Thomson effect in a nonmagnetic conductor [61] (Fig. 5). By replacing the lock-in source with a periodically modulated heat current and microwave, lock-in thermography measurements of the thermal Hall effect [62] and unidirectional spin-wave heat conveyer [55,57] have also been realized, respectively. In addition to infrared-camera-based methods, emerging techniques such as lock-in thermoreflectance [63] and lock-in scanning thermal microscopy [64,65] have also been employed for observing the magneto-thermoelectric and thermo-spin effects in recent years. As introduced below, various unobserved phenomena remain in spin caloritronics, and further development of thermal measurement techniques is desired to pioneer new physics and functionalities.

Thermal conductivity measurements for spintronic devices and magnetic multilayer films are also being conducted. Such measurements have been mainly performed by means of the time-domain thermoreflectance (TDTR) method [23,24,27]. TDTR is a pump-probe technique for evaluating thermal transport properties of thin films and interfaces. In TDTR, ultrafast femto- or pico-second pulsed lasers are divided into pump and probe beams. The pump beam periodically heats the sample surface, while the delayed probe beam detects the resulting change in optical reflectance, which is proportional to the transient surface temperature. By analyzing the time-dependent thermoreflectance response with a thermal diffusion model, TDTR can quantify the thermal conductivity over a wide dynamic range, from ~1000 $\mathrm{Wm^{-1}K^{-1}}$ [66] to ~0.01 $\mathrm{Wm^{-1}K^{-1}}$ [67,68]. In addition to thermal conductivity, TDTR is also widely used to evaluate interfacial thermal resistance and heat capacity (note that the data in Fig. 4 were obtained by the TDTR method). A key advantage of TDTR is its simple sample preparation; the sample surface only needs to be coated with a thin metal transducer layer, which serves as both the optical absorber and detector. Therefore, it requires minimal sample preparation and no microfabrication design, such as electrical heaters or temperature sensors. Urgent challenges for thermal transport in spin caloritronics include elucidating the origin of giant MTR in the Cu/CoFe multilayers [23,24]. Clues to unravel it are gradually being obtained [27,69].

### *3.2. Fundamental physics*

#### *3.2.1. Transverse thermo-spin effects*

All the thermo-spin effects introduced in Section 2.3 are the longitudinal effects where heat and spin currents are interconverted in the same direction. In addition to these effects, research on transverse thermo-spin effects is also gradually progressing. Multiple research groups reported the observation of the spin Nernst effect, which is the thermal counterpart to the spin Hall effect where a spin current is generated perpendicular to an applied temperature gradient via the spin-orbit interaction in nonmagnetic materials, around the same time in 2017 [70-72]. While these experiments also suggest the existence of its reciprocal phenomenon, the spin Ettingshausen effect, direct observation of the transverse heat current induced by the spin current injected into nonmagnetic materials has yet to be reported (Fig. 6). Compared to the longitudinal thermo-spin effects, the transverse thermo-spin effects have been verified only in a limited number of materials, leaving many research challenges.

#### *3.2.2. Thomson effects in spin caloritronics*

Spin caloritronics has achieved success by incorporating spin transport properties into thermal transport and thermoelectric effects or by exploring spin-current versions of thermoelectric conversion phenomena. However, the Thomson effect, which is one of the fundamental thermoelectric effects, had not appeared in spin caloritronics (Fig. 6). The conventional Thomson effect is a phenomenon where heat absorption or release occurs when a charge current flows through a conductor under a temperature gradient. Here, the directions of the charge and heat currents are parallel to each other. Unlike the Peltier effect, the heat absorption or release induced by the Thomson effect is proportional to both the applied charge current and temperature difference and occurs even without a junction between different materials. The performance of the Thomson effect is expressed by the Thomson coefficient $\tau$, which follows the Thomson (or Kelvin) relation $\tau = T\,\mathrm{d}S/\mathrm{d}T$ with $S$ and $T$ respectively being the Seebeck coefficient and absolute temperature. That is, the Thomson effect manifests in materials with a large temperature dependence of the Seebeck coefficient, which is beyond the linear response regime. It should be noted that all the spin-caloritronic phenomena introduced thus far are the transport phenomena in the linear response regime; even if a temperature gradient is applied to a sample, it is assumed that the transport coefficients within the sample remain constant. However, starting around 2020, spin-caloritronic phenomena in the nonlinear regime, where the temperature dependence of the transport coefficients plays an essential role, began to be explored.

As mentioned above, the influence of magnetic fields and magnetism on the Seebeck and Peltier effects, i.e., the magneto-Seebeck and Peltier effects, has been studied for many years. However, before the advent of spin caloritronics, research on how the Thomson effect depends on magnetic fields and magnetism had not progressed due partially to the measurement difficulty. Amidst this situation, in 2020, Uchida et al. established a method using the lock-in thermography to precisely measure the heat absorption and release occurring when a charge current flows through a conductor under a temperature gradient and magnetic field [59]. The measurements using a $Bi_{88}Sb_{12}$ alloy, known to exhibit the large magneto-Seebeck effect, revealed that temperature modulation signals in the alloy was proportional to both the charge current and temperature difference and their magnitude increased significantly with an applied magnetic field. This represents the first direct observation of the magnetic field dependence of the Thomson effect in a non-magnetic material, i.e., the magneto-Thomson effect. In 2023, by applying the same measurement technique to a ferromagnetic $Ni_{95}Pt_5$ alloy, Modak et al. reported the direct observation of the anisotropic magneto-Thomson effect, where the Thomson coefficient changes anisotropically depending on the magnetization direction [60]. The (anisotropic) magneto-Thomson effect is classified as a longitudinal magneto-thermoelectric effect because the applied charge current and temperature gradient are parallel. In 2025, Takahagi et al. achieved the observation of the transverse Thomson effect, a phenomenon where heat absorption or release occurs when a heat current, charge current, and magnetic field are applied to a conductor in orthogonal directions [61]. Unlike the Thomson relation for the longitudinal effects, the transverse Thomson coefficient depends not only on the temperature derivative of the Nernst coefficient but also on its absolute value. These results mark the beginning of the unexplored field of nonlinear spin caloritronics.

However, the spin-current version of the Thomson effect remains entirely unexplored. At least, two types of such phenomena should exist: a spin-dependent Thomson effect originating from a conduction-electron spin current and a spin Thomson effect originating from a magnon spin current (Fig. 6). These phenomena can be defined as the generation of heat absorption or release proportional to both the spin current and temperature gradient in magnetic materials or hybrid structures, but suitable material systems or measurement methods for observing the spin and spin-dependent Thomson effects have not yet been established.

The exploration of spin caloritronics physics beyond the linear response regime is still in its infancy. Future research is expected to systematize nonlinear transport physics based on heat-charge-spin interactions. Note that the term "nonlinear" used for the discussions of the Thomson effects is to describe outputs proportional to both an applied charge or spin current and temperature gradient, originating from the temperature dependence of linear-response transport coefficients. Beyond this framework, in 2024, Arisawa et al. observed a nonlinear Nernst effect, where the transverse thermopower is proportional to the square of the applied longitudinal temperature gradient, in a junction structure consisting of a superconducting MoGe thin film and YIG [73]. Subsequently, Hirata et al. reported the observation of a nonlinear Seebeck effect, where the longitudinal thermopower is proportional to the square of the temperature gradient, in a $Ni_{81}Fe_{19}$/Pt bilayer structure [74]. While the mechanisms of these phenomena are entirely different from the Thomson-effect-related phenomena discussed above, they hold potential for new developments in thermoelectrics and spin caloritronics.

*3.2.3. Polarization transport in ferroelectrics*

Efforts to apply physics developed in spin caloritronics to different material systems are also progressing. As symbolized by the history of spin caloritronics, the science of transport phenomena advances by introducing new transport carriers and their driving forces. In 2021, Bauer et al. theoretically proposed that a nonequilibrium flow of electric polarization in ferroelectrics can induce thermal conduction and thermoelectric conversion [75]. Since ferroelectrics are insulators, the conventional Seebeck and Peltier effects do not manifest. Nevertheless, through the collective excitation of electric polarization, called ferrons, electric polarization transport phenomena can exist. In a similar manner to the magnon spin current that is a net flow of magnetic dipoles, a ferron current is a net flow of electric dipoles without accompanying a charge current. It is driven not by an electric field but by an electric field gradient, making it fundamentally different from a displacement current. Although thermoelectric effects due to ferron transport have not been experimentally observed yet, the thermal conduction by ferrons has been demonstrated through measurements of the electric field dependence of thermal conductivity in ferroelectrics. In 2023, Wooten et al. systematically investigated the electric field dependence of the thermal conductivity, thermal diffusivity, and longitudinal sound velocity in ferroelectric lead zirconate titanate [76]. They revealed that these physical quantities exhibited distinct changes depending on the electric polarization direction. The observed changes were quantitatively reproduced by a theoretical model combining piezoelectric strain and phonon anharmonicity. This demonstrated that ferrons vibrating at optical phonon frequencies in the THz range can be converted into the acoustic phonon regime, which acts as a heat carrier, via piezoelectric strain and phonon anharmonicity. While ferrons have primarily been treated as phonons capable of transporting net electric polarization in this experiment, Adachi et al. constructed a ferron transport model for electronic ferroelectrics, beginning to explore the physics of transport phenomena driven by non-phononic ferrons [77]. These efforts exemplify how spin caloritronics has created novel physics and functionalities in entirely different material systems. Recently, the nonlocal ferron drag effect, a thermoelectric conversion phenomenon emerging in a metal formed on a ferroelectric, has also been proposed theoretically [78]. Experimental verification of this phenomenon, which serves as an interface between ferron transport and electronics, is highly desired.

*3.3. Materials science and engineering applications*

For the continued advancement of spin caloritronics, it is crucial not only to pioneer new physics and principles but also to develop applications by improving the performance of the magneto-thermoelectric, thermomagnetic, and thermo-spin effects. Conventional spin caloritronics research has primarily utilized model materials with known fundamental properties to elucidate the heat-spin current conversion mechanisms. Now that the fundamental physics of spin caloritronics in the linear response regime has been mostly elucidated, the time is ripe to advance materials science and device applications to realize thermal engineering technologies based on spin caloritronics.

One significance of applying spin caloritronics to thermal engineering lies in its ability to realize transverse thermoelectric conversion, where heat and charge currents are converted in orthogonal directions [79,80]. The Seebeck effect, which drives existing longitudinal thermoelectric devices, necessitates complex module structures with numerous junctions comprising different materials. Consequently, energy conversion loss at the junction interfaces, high manufacturing cost, low durability, and limited design freedom have hindered practical applications. In contrast, utilizing the transverse thermoelectric effects allows power generation and thermal energy harvesting simply by spreading materials along heat sources. This enables the construction of thermoelectric modules with no junctions, potentially resolving the fundamental issues faced by the Seebeck-effect-based devices. However, the transverse thermopower, i.e., a transverse electric field divided by an applied longitudinal temperature gradient, due to ANE or the inverse spin Hall effect induced by SSE is more than one order of magnitude lower than the Seebeck coefficient of thermoelectric materials, necessitating a breakthrough for practical applications.

As mentioned in Section 2.1, with the advancement of topological materials science, research on ANE has accelerated rapidly in recent years. The anomalous Nernst coefficient, transverse thermopower due to ANE, one order of magnitude larger than that of typical ferromagnets like Fe or Ni has been observed in various materials. It is important to note that the anomalous Nernst coefficient generally does not correlate with the magnitude of saturation magnetization; magnetic materials with large magnetization are not necessarily excellent ANE materials [11]. In fact, topological materials can exhibit large ANE even with small saturation magnetization [12,14,15]. However, even in topological materials, the anomalous Nernst coefficient does not reach 10 $\mu VK^{-1}$ at room temperature, making applications of ANE as energy harvesting devices requiring large output power difficult. Due to the small anomalous Nernst coefficient and high thermal conductivity of known magnetic materials, the figure of merit for ANE is still in the order of $10^{-3}$ or less around room temperature. A promising application candidate for ANE is a heat flux sensor. Even with low output power, incorporating a thermopile structure to enhance an open-circuit voltage enables operation as a highly sensitive heat flux sensor [21,22]. A problem with a conventional heat flux sensor driven by the Seebeck effect is that the high thermal resistance of the sensor itself disturbs the heat flux distribution to be measured. The advantage of ANE lies in its ability to reduce the thickness of materials along the temperature gradient direction, thereby dramatically reducing the thermal resistance of the sensor.

Efforts to enhance the performance of transverse thermoelectric conversion are gaining momentum, utilizing not only homogeneous single materials but also hybrid structures or composite materials made of dissimilar materials and nanoscale precipitates. Homogeneous materials do not necessarily exhibit superior functionalities, which is evident from the examples of permanent magnets and soft magnetic materials in practical use. Also in spin caloritronics, it has been demonstrated that short thermal annealing of iron-based amorphous alloys significantly enhances the anomalous Nernst coefficient without altering the material's average composition [81]. Furthermore, in 2021, Zhou et al. proposed and demonstrated a novel transverse thermoelectric conversion process emerging in hybrid materials composed of magnetic metals and thermoelectric semiconductors [82]. This process, termed the Seebeck-driven transverse magneto-thermoelectric generation or Seebeck-effect-driven anomalous Hall effect, arises because the Seebeck effect in the thermoelectric semiconductor induces a charge current in the magnetic metal, thereby generating the anomalous Hall effect contributing to the transverse thermopower [Fig. 7(a)]. The transverse thermopower of the Seebeck-effect-driven anomalous Hall effect can be designed with various degrees of freedom, such as the combination of magnetic metals and thermoelectric semiconductors and their size ratio. Therefore, it is possible to achieve the transverse thermopower and figure of merit far exceeding those of ANE alone through optimization [Fig. 7(b)] [82,83]. Recently, the nonmagnetic version of this process, the Seebeck-effect-driven ordinary Hall effect, has also been observed in bulk composites, confirming its versatility [84]. However, the Seebeck-effect-driven ordinary and anomalous Hall effects are not universally useful and face many challenges, such as a trade-off between transverse thermopower and output power.

The thermoelectric conversion based on SSE also requires a junction between different materials. By combining SSE in insulators and the inverse spin Hall effect in metals, it is possible to achieve thermoelectric generation driven by heat in insulators, which was impossible with conventional thermoelectric effects [39,40,85]. However, since the thermoelectric generation based on SSE occurs only within a spin or magnon diffusion length scale in the vicinity of the insulator/metal interface, the majority of the device far from the interface cannot contribute to the energy conversion. Consequently, energy device applications of SSE remain challenging as long as conventional thin-film bilayer structures are used. As a solution, utilizing bulk nanocomposites, aggregates of nanoscale precipitates and interfaces, has been proposed. It has been demonstrated that nanocomposites allow SSE to contribute to thermoelectric generation even in macroscale materials, although their microstructures are yet to be optimized [86,87].

Throughout the history of spin caloritronics, it was crucial to clearly separate various magneto-thermoelectric and thermo-spin conversion responses to elucidate their mechanisms and behaviors. However, from the perspective of thermoelectric applications, there is no need to separate different phenomena. Rather, it is more rational to pursue higher output through hybrid power generation or cooling combining multiple phenomena. While the performance of each individual transverse thermoelectric conversion phenomenon is inferior to the Seebeck or Peltier effect, the synergistic effect of multiple phenomena holds the potential to achieve high-output transverse thermoelectric conversion unattainable by any single phenomenon alone. The hybrid transverse magneto-thermoelectric conversion was first observed in junction structures comprising a ferromagnetic metal and ferrimagnetic insulator, where SSE and ANE occur simultaneously when the ferromagnetic metal exhibits the inverse spin Hall effect [88]. However, its impact was limited because both SSE and ANE exhibit small transverse thermopower. The systems showing the Seebeck-effect-driven anomalous Hall effect [Fig. 7(a)] also exhibit hybrid transverse magneto-thermoelectric conversion due to the simultaneous occurrence of ANE.

High-output hybrid transverse magneto-thermoelectric conversion, suitable for energy harvesting and cooling applications, was reported by Uchida et al. in 2024 through experiments using artificially tilted multilayers (ATMLs) showing a large off-diagonal Seebeck or Peltier effect, which is a transverse thermoelectric effect due to a macroscale anisotropic structure [89]. In this experiment, they prepared ATML by alternately stacking $Bi_{88}Sb_{12}$ slabs, which exhibit the large magnetic-field-dependent magneto-thermoelectric effects, and $Bi_{0.2}Sb_{1.8}Te_3$ slabs, which exhibit the large field-independent Peltier effect. In this structure, applying an external magnetic field significantly enhances cooling performance via transverse thermoelectric conversion, achieved through a hybrid action involving three effects: the off-diagonal Peltier effect, magneto-Peltier effect, and ordinary Ettingshausen effect. This concept is effective not only for cooling but also for thermoelectric generation; hybridization of the off-diagonal Seebeck effect and ANE has also been demonstrated in $Co_2MnGa/Bi_{0.2}Sb_{1.8}Te_3$ and $Co_2MnGa/Bi_2Te_3$ ATMLs [Figs. 7(c) and (d)] [90]. The magnetic-field-induced or magnetization-dependent modulation of the figure of merit in such ATMLs is significantly greater than the figure of merit of the Nernst effects alone. This results from the synergistic effect with the off-diagonal Seebeck effect; the greater the thermopower of the off-diagonal Seebeck effect, the greater the modulation of the figure of merit due to the coexisting magneto-thermoelectric effects [see Fig. 7(e) and Ref. [90] for details]. These efforts have dramatically increased the output of transverse thermoelectric conversion, leading to the development of transverse thermoelectric conversion modules that exhibit a power generation density larger than commercial thermoelectric modules, where $SmCo_5/Bi_{0.2}Sb_{1.8}Te_3$ ATMLs show record-high performance around room temperature owing to their tiny interfacial electrical and thermal resistances [91,92]. The figure of merit in $SmCo_5/Bi_{0.2}Sb_{1.8}Te_3$ ATML at room temperature reaches ~0.3, which is more than two orders of magnitude higher than that for ANE. This value is promising, considering that the performance of longitudinal thermoelectric conversion degrades by several tens of percent due to modularization [80]. In the high-temperature range, the figure of merit of 0.7 (at 980 K) due to the off-diagonal Seebeck effect has been reported for the goniopolar material $Re_4Si_7$ [93]. At present, the only transverse thermoelectric conversion that can contribute to power generation (cooling) applications requiring high figure of merit is the off-diagonal Seebeck (Peltier) effect. However, the integration of spin-caloritronic principles, materials, and techniques holds the potential to further boost such application research, as illustrated in the examples above.

## 4. Conclusion

In this review, we have provided an overview of the history of spin caloritronics and research on various transport phenomena arising from heat-charge-spin current interactions. We have also outlined future prospects for this field from the perspectives of thermal measurement techniques, condensed matter physics, materials science, and engineering applications. We anticipate that continuously incorporating insights, technologies, and materials from diverse research fields will lead to the discovery of further new principles and phenomena, as well as improvements in magneto-thermoelectric and thermo-spin conversion efficiencies. Many spin-caloritronic phenomena, such as SSE, SPE, and the Seebeck-effect-driven anomalous Hall effect, require interfaces and junctions of different materials, optimizations of which are crucial for further development and practical applications of spin caloritronics. Regarding applications, this review primarily focused on transverse thermoelectric conversion and details on thermal switching using MTR were omitted. However, if giant MTR, currently realized only in nanoscale multilayer films, could be achieved in macroscale materials, it would represent a significant step toward realizing active thermal management technologies.

## Declaration of Competing Interest

The authors declare that they have no known competing financial interests or personal relationships that could have appeared to influence the work reported in this paper.

## Acknowledgements

The authors thank G. E. W. Bauer and E. Saitoh for valuable discussions. The previously reported experimental results introduced in this review (Figs. 4 and 7) were obtained from the collaboration with F. Ando, S. Biswas, D. Chiba, R. Guo, R. Iguchi, V. K. Kushwaha, F. Makino, A. Miura, Y. Miura, R. Modak, T. Morita, T. Ohkubo, Y. Sakuraba, H. Sepehri-Amin, J. Shiomi, J. Uzuhashi, B.

Xu, T. Yagi, K. Yamamoto, Y. Yamashita, and W. Zhou. The authors were mainly supported by ERATO "Magnetic Thermal Management Materials" (grant no. JPMJER2201) from Japan Science and Technology Agency and Grant-in-Aid for Scientific Research (S) (grant no. 22H04965) from Japan Society for the Promotion of Science.

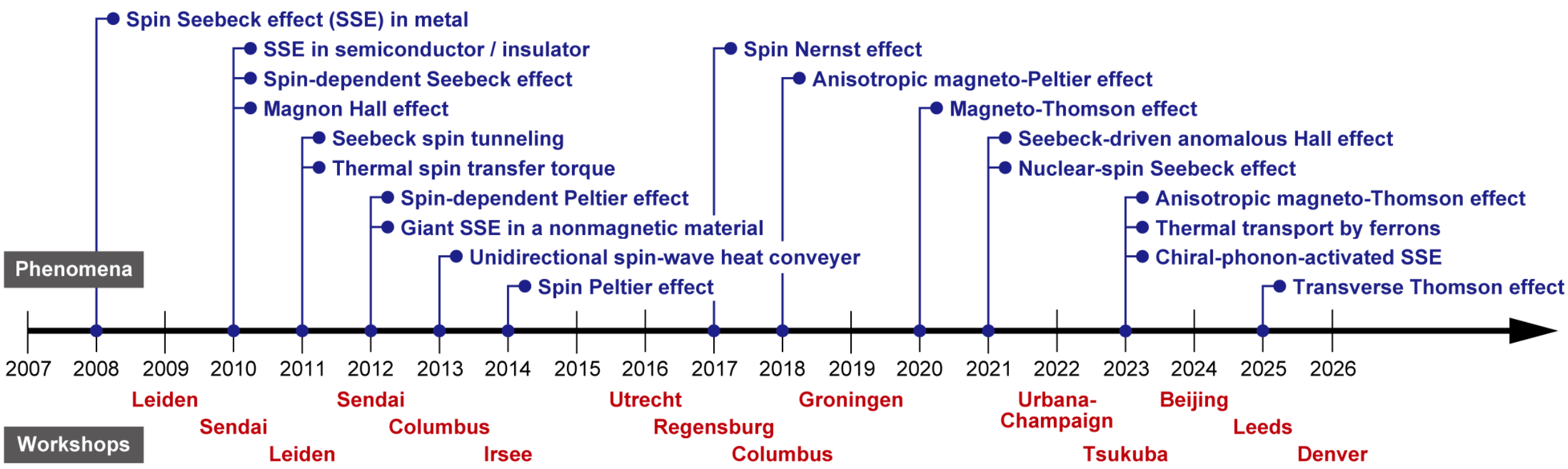

**Fig. 1.** Physical phenomena discovered in spin caloritronics and cities hosting the International Workshop on Spin Caloritronics.

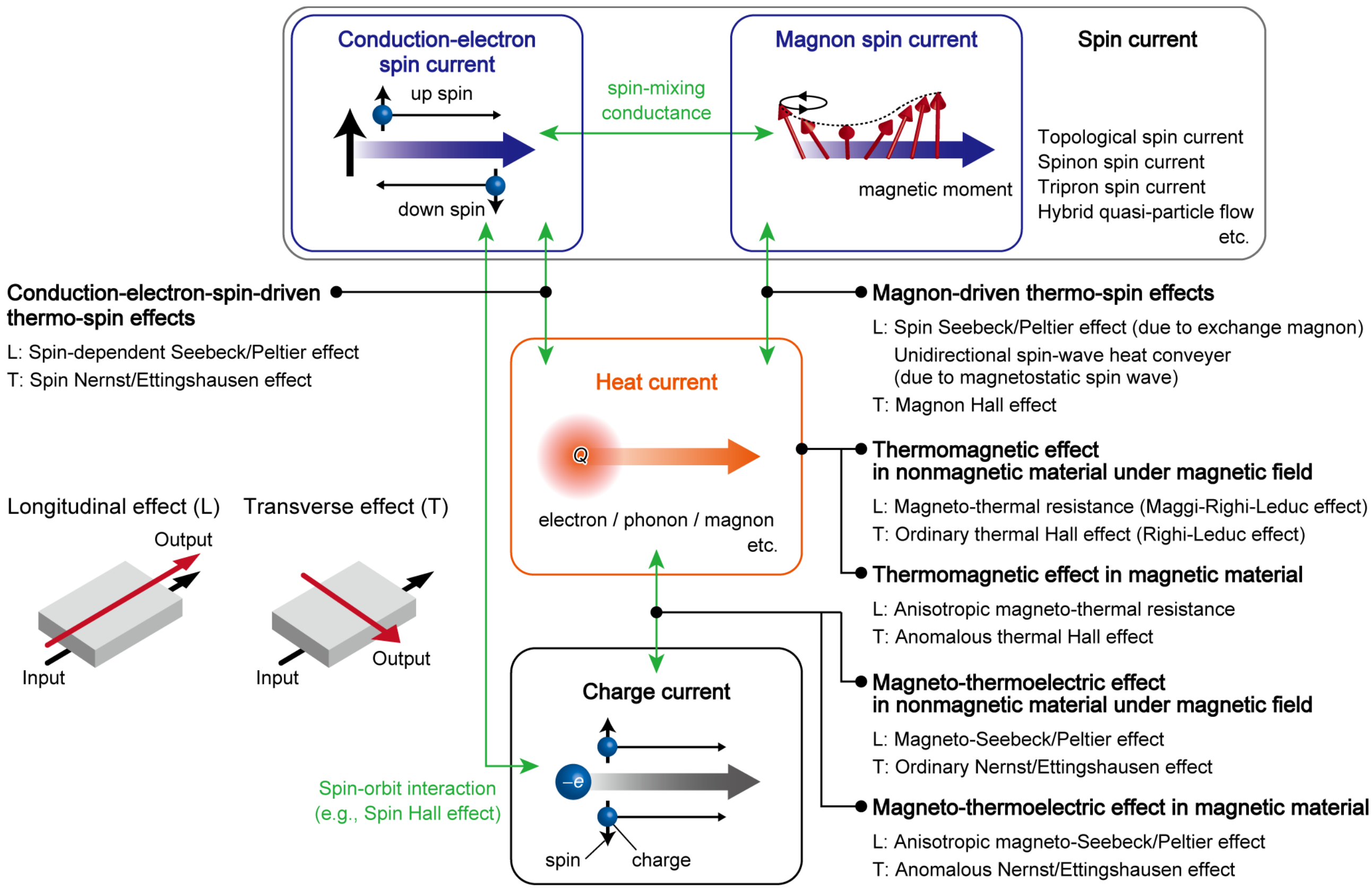

**Fig. 2.** Conversion phenomena between the charge current, heat current, conduction-electron spin current, and magnon spin current and their categorization. $-e$ and $Q$ denote the electron charge and heat, respectively.

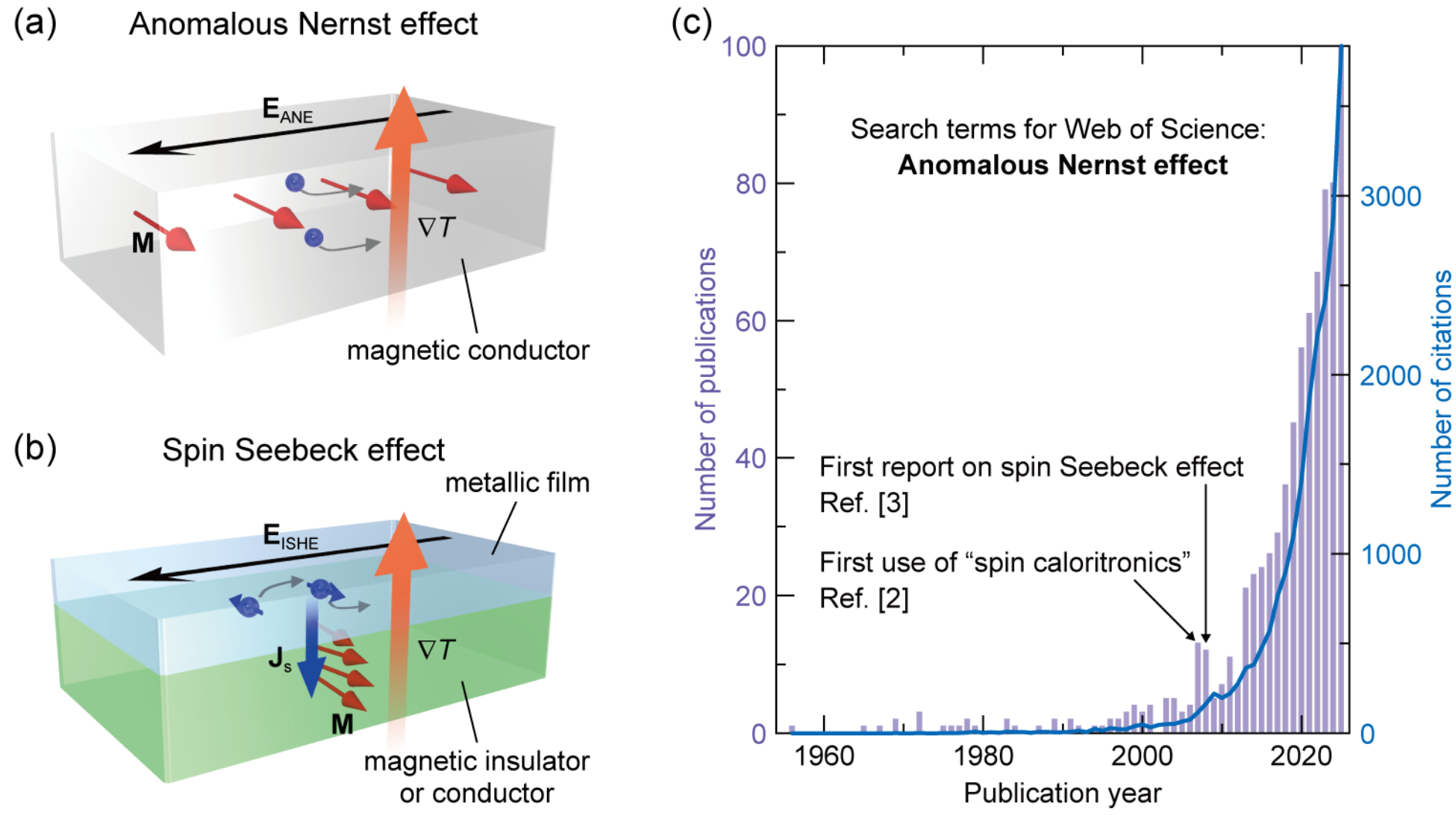

**Fig. 3.** (a) Schematic of ANE in a magnetic conductor. (b) Schematic of the inverse spin Hall effect induced by SSE in a junction structure comprising a metallic film and magnetic insulator or conductor. $\nabla T$, $\mathbf{M}$, $\mathbf{E}_{\mathrm{ANE(ISHE)}}$, and $\mathbf{J}_{\mathrm{s}}$ denote the temperature gradient, magnetization vector, electric field induced by ANE (inverse spin Hall effect), and spatial direction of a spin current, respectively. (c) Numbers of publications and their citations from 1956 to 2025 for search terms "Anomalous Nernst effect" entered into Web of Science.

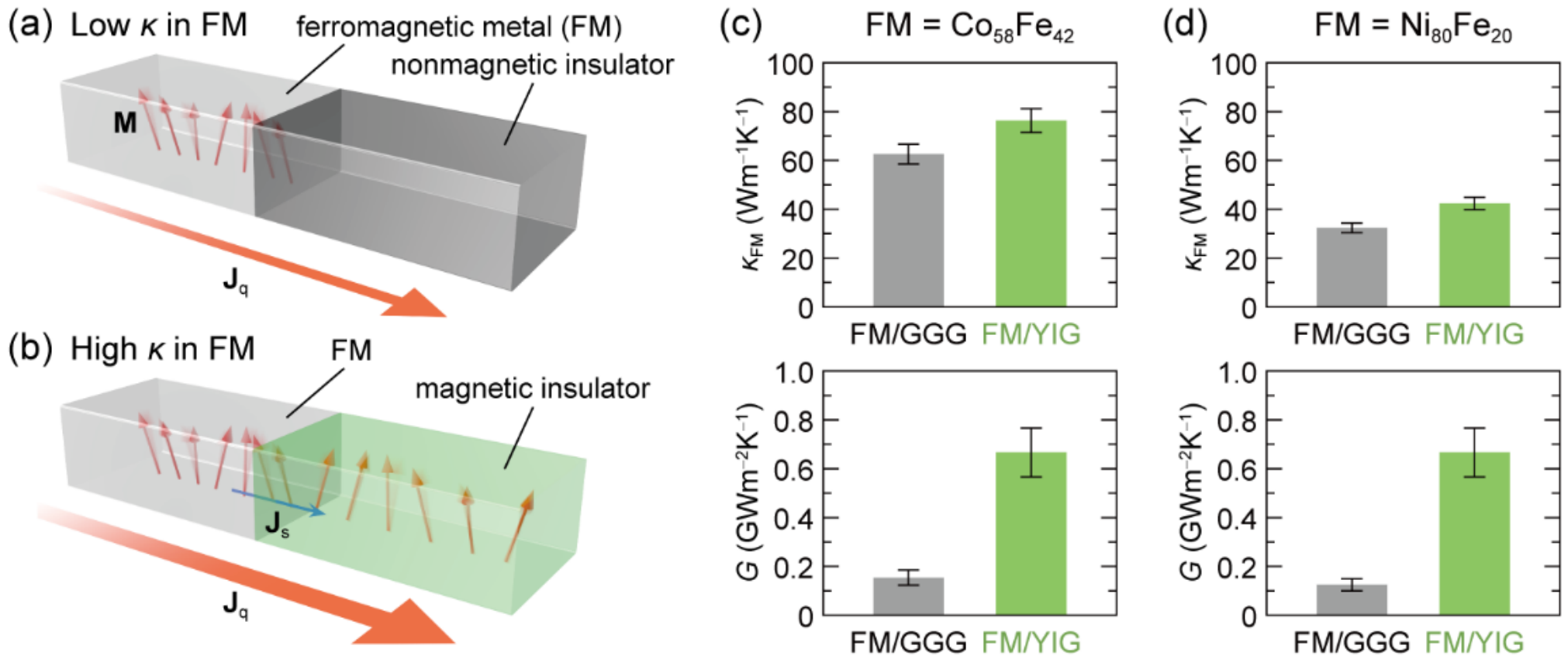

**Fig. 4.** (a),(b) Schematic of thermal transport engineerring in ferromagnetic metal (FM)/insulator junctions via nonequilibrium magnon spin currents. Depending on the boundary condition for $\mathbf{J}_{\mathrm{s}}$, the thermal conductivity $\kappa$ of FM ($\kappa_{\mathrm{FM}}$) in a FM/nonmagnetic insulator junction (a) differs from that in a FM/magnetic insulator junction (b). $\mathbf{J}_{\mathrm{q}}$ denotes the heat current. (c),(d) $\kappa_{\mathrm{FM}}$ and interfacial thermal conductance $G$ in $Co_{58}Fe_{42}$/insulator (c) and $Ni_{80}Fe_{20}$/insulator (d) junctions [27]. Here, GGG and YIG are used as nonmagnetic and magnetic insulators, respectively.

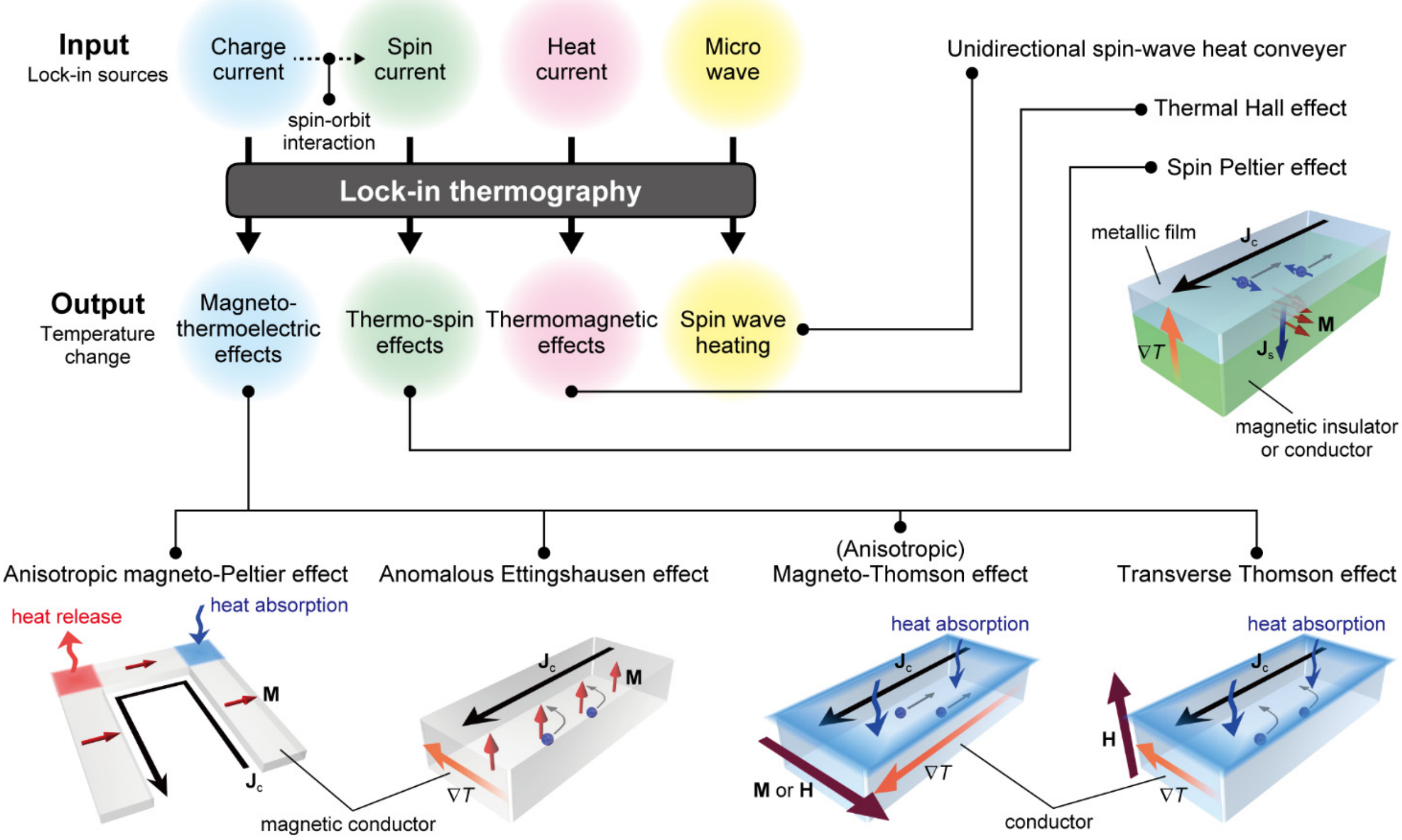

**Fig. 5.** Lock-in thermography as a tool for investigating spin caloritronics. Spin-caloritronic phenomena that outputs a heat current or heat absorption/release can be measured by the lock-in thermography method. $\mathbf{J}_c$ and $\mathbf{H}$ denote the charge current and magnetic field vector, respectively.

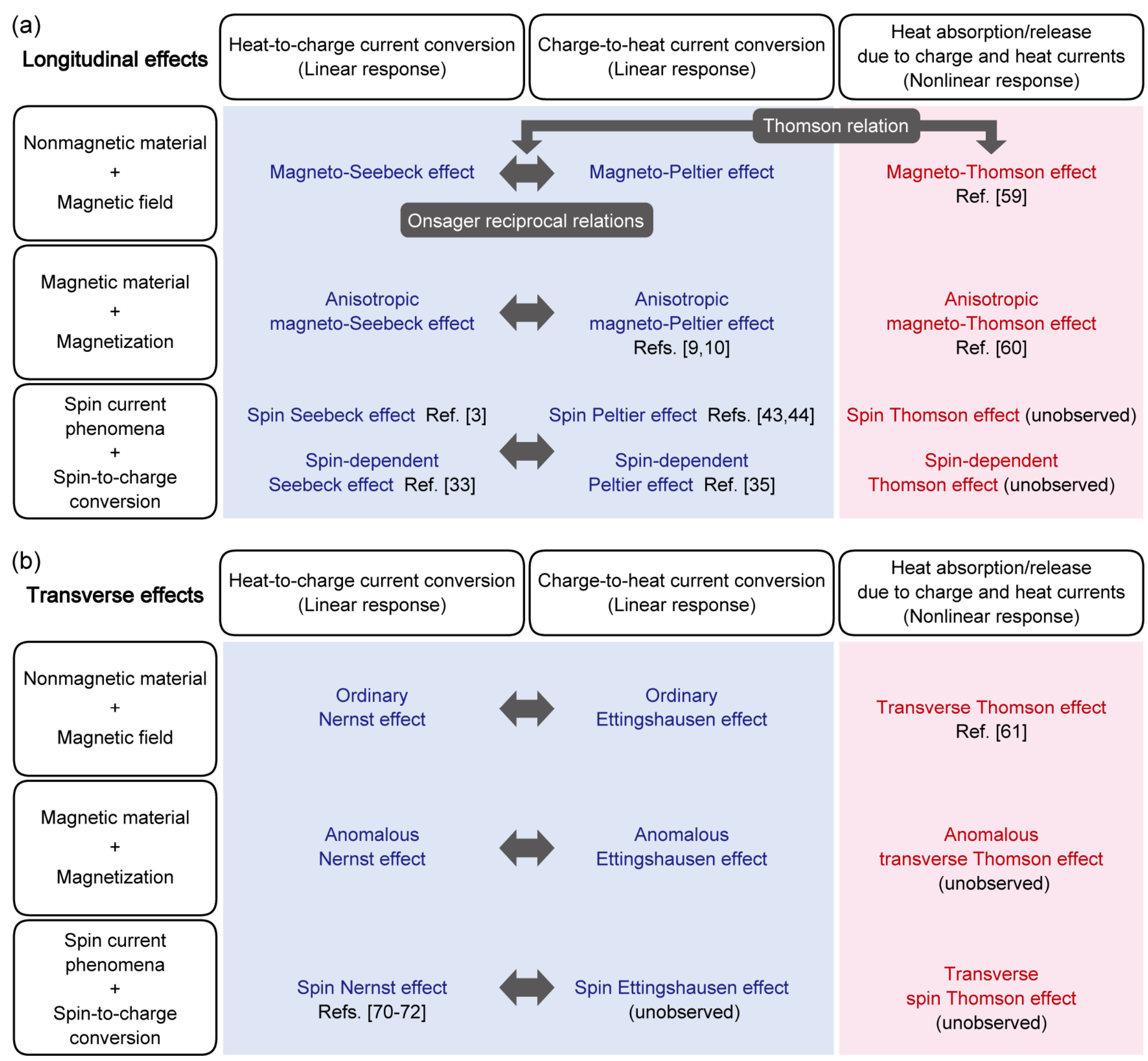

**Fig. 6.** Linear response magneto-thermoelectric, thermo-spin, and Thomson effects in spin caloritronics for the longitudinal effects (a) and the transverse effects (b).

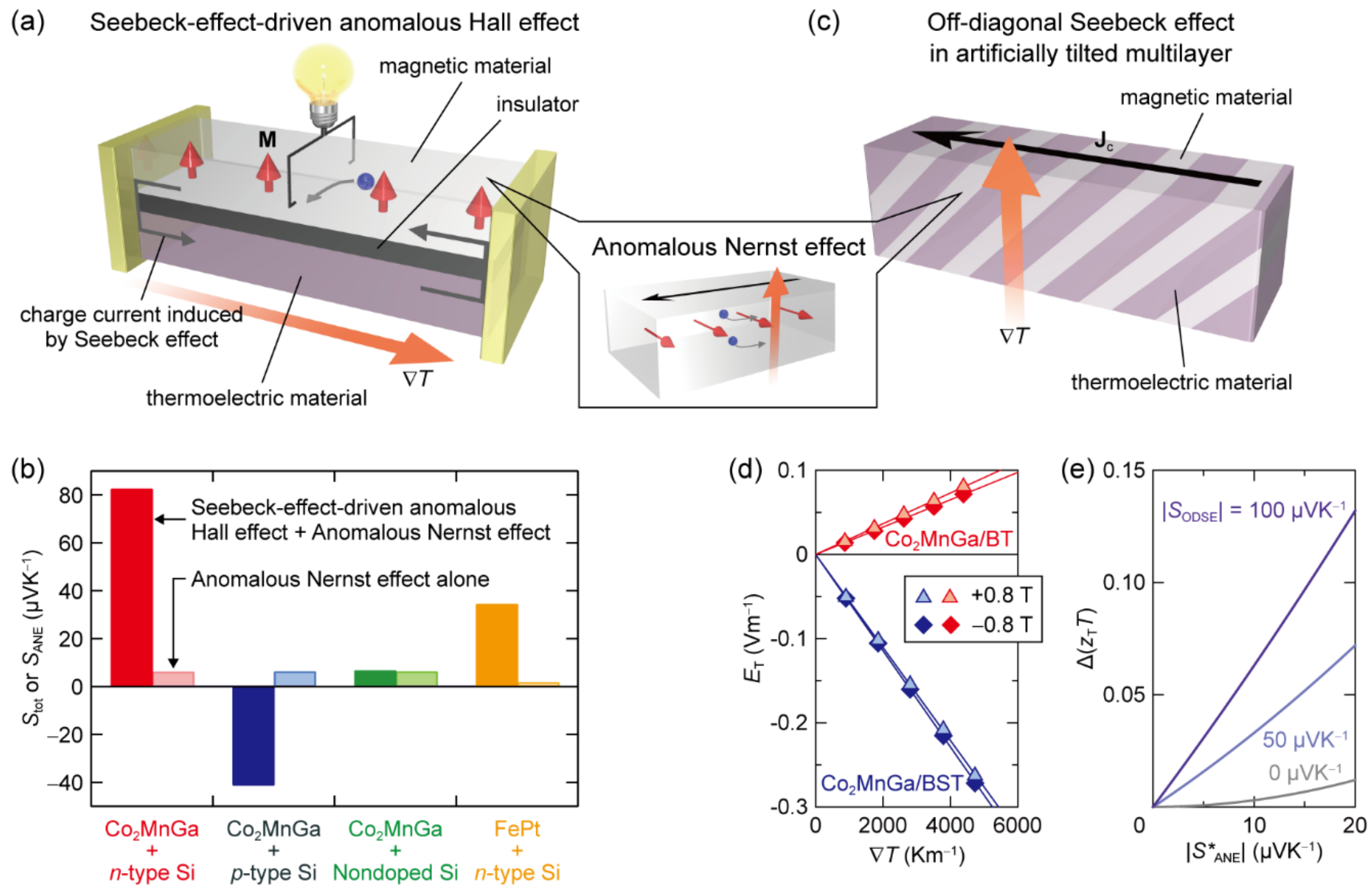

**Fig. 7.** (a) Schematic of the Seebeck-effect-driven anomalous Hall effect (Seebeck-driven transverse magneto-thermoelectric generation) in the closed-loop junction comprising magnetic and thermoelectric materials. (b) Measured transverse thermopower in the closed-loop junctions comprising a ferromagnetic metal film ($Co_2MnGa$ or FePt) and Si substrate [82]. $S_{tot}$ and $S_{ANE}$ denote the total transverse thermopower in the closed-loop junction and the anomalous Nernst coefficient of the ferromagnetic metal, respectively. (c) Schematic of the off-diagonal Seebeck effect in ATML comprising magnetic and thermoelectric materials. In the magnetic materials, ANE also occurs. (d) $\nabla T$ dependence of the transverse electric field $E_T$ in $Co_2MnGa/Bi_{0.2}Sb_{1.8}Te_3$ (BST) and $Co_2MnGa/Bi_2Te_3$ (BT) ATMLs at a magnetic field of ±0.8 T [90]. (e) Simulated ANE-induced modulation of the dimensionless figure of merit for the transverse thermoelectric generation $z_TT$ with the hybridized off-diagonal Seebeck effect in ATML [90]. $S_{ODSE}$ and $S^*_{ANE}$ denote the transverse thermopower due to the off-diagonal Seebeck effect and the effective anomalous Nernst coefficient considering a shunting effect in ATML, respectively.